\documentstyle[12pt]{article}  

\def\R{ {\rm R \kern -.31cm I \kern .15cm}}
\def\C{ {\rm C \kern -.15cm \vrule width.5pt \kern .12cm}}
\def\Z{ {\rm Z \kern -.27cm \angle \kern .02cm}}
\def\N{ {\rm N \kern -.26cm \vrule width.4pt \kern .10cm}}
\def\1{{\rm 1\mskip-4.5mu l} }
\def\lsim{\raise0.3ex\hbox{$<$\kern-0.75em\raise-1.1ex\hbox{$\sim$}}}
\def\gsim{\raise0.3ex\hbox{$>$\kern-0.75em\raise-1.1ex\hbox{$\sim$}}}
\def\noi{\noindent}

\def\beq{\begin{equation}}   \def\eeq{\end{equation}}
\def\bea{\begin{eqnarray}}  \def\eea{\end{eqnarray}}
\def\nn{\nonumber}
\def\noi{\noindent}
\def\beeq{\begin{eqnarray}} \def\eeeq{\end{eqnarray}}
\newcommand\mysection{\setcounter{equation}{0}\section}
\renewcommand{\theequation}{\thesection.\arabic{equation}}
\newcounter{hran} \renewcommand{\thehran}{\thesection.\arabic{hran}}

\def\bmini{\setcounter{hran}{\value{equation}}
  \refstepcounter{hran}\setcounter{equation}{0}
  \renewcommand{\theequation}{\thehran\alph{equation}}\begin{eqnarray}}

\def\bminiG#1{\setcounter{hran}{\value{equation}}
\refstepcounter{hran}\setcounter{equation}{-1}
\renewcommand{\theequation}{\thehran\alph{equation}}
\refstepcounter{equation}\label{#1}\begin{eqnarray}}

\def\emini{\end{eqnarray}\relax\setcounter{equation}{\value{hran}}\renewcommand{\theequation}{\thesection.\arabic{equation}}}

\begin{document} 
\centerline{\Large\bf Cosmological evolution in compactified 
Ho{\v r}ava-Witten} 
\vskip 3 truemm \centerline{\Large\bf  theory induced by matter on the
branes} 
\vskip 1 truecm

\centerline{\bf Ulrich Ellwanger}
\centerline{Laboratoire de Physique Th\'eorique\footnote{Unit\'e Mixte de
Recherche - CNRS - UMR 8627}}  \centerline{Universit\'e de Paris XI, B\^atiment
210, F-91405 ORSAY Cedex, France}
\vskip 2 truecm

\begin{abstract}
The combined Einstein equations and scalar equation of motion in the
Ho{\v r}ava-Witten scenario of the strongly coupled heterotic string
com\-pac\-ti\-fied on a Calabi-Yau manifold are solved in the presence
of additional matter densities on the branes. We take into account the
universal Calabi-Yau modulus $\varphi$ with potentials in the 5-$d$
bulk and on the 3-branes, and allow for an arbitrary coupling of the
additional matter to $\varphi$ and an arbitrary equation of state. No
ad hoc stabilization of the five dimensional radius is assumed. The
matter densities are assumed to be small compared to the potential for
$\varphi$ on the branes; in this approximation we find solutions in the
bulk which are exact in y and t. Depending on the coupling of the
matter to $\varphi$ and its equation of state, various solutions for
the metric on the branes and in the 5-$d$ bulk are obtained: Solutions
cor\-res\-pon\-ding to a ``rolling radius'', and solutions with a
static 5-$d$ radius, which reproduce the standard cosmological
evolution.
\end{abstract} 

\vskip 1.5 truecm
\noi LPT Orsay 00-02 \par
\noi January 2000 \par

\newpage
\pagestyle{plain}
\baselineskip 18pt

\mysection{Introduction} \hspace*{\parindent} Motivated, to a large
extent, by the Ho{\v r}ava-Witten scenario of the strongly coupled
heterotic string \cite{hw95}, the cosmology of a five dimensional
universe bounded by 3-branes has recently been the subject of many
investigations. If one estimates the parameters in the Ho{\v
r}ava-Witten scenario using, as input, the phe\-no\-me\-no\-lo\-gi\-cal
values of Newton's constant, $M_{GUT}$ and $\alpha_{GUT}$
\cite{w96,bd96} one finds that there exists a regime below $M_{GUT}$
(the inverse size of a Calabi-Yau manifold, on which 6 dimensions are
compactified) in which a five dimensional description of our universe
is appropriate. \par

The topology of the five dimensional universe is the one of a $S^1/Z_2$
orbifold: There are two distinct 3-branes, and the fields in the bulk
between the branes are either symmetric or anti-symmetric with respect
to $Z_2$, i.e. reflections at the branes. The field content in the five
dimensional bulk is the one of $N = 1$ (gauged) supergravity, plus
``matter'' originating from a 3-form in 11 dimensions, and from
internal degrees of freedom of the metric (moduli)
[4 -- 7]. On the branes we have, in
addition, matter originating from $E_8$ Yang-Mills theories in 10
dimensions. \par

Solutions to the equations of motion in the five dimensional bulk have
been obtained in 
[4 -- 12].
Due to the presence of potentials for the moduli on the branes (with
opposite signs) one finds that the moduli configurations vary with $y$,
the fifth coordinate, leading to a $y$-dependent energy-momentum tensor
in the bulk. \par

Independently from the Ho{\v r}ava-Witten theory, scenarios for the
cosmology of a five dimensional universe with branes have been
developed. Assuming a $y$-indepen\-dent cosmological constant in the
bulk, and fine-tuned cosmological constants on the branes (with
opposite signs), Randall and Sundrum \cite{rs99a} (RSI) found a static
solution of the Einstein equations in the bulk with an exponential
dependence of the metric on $y$. They argued that this phenomenon could
solve the hierachy problem, provided our four dimensional universe
lives on a brane with negative tension (negative cosmological
constant). Alternatively, if we live on a brane with positive tension
surrounded by a bulk with a $y$-independent cosmological constant
\cite{g98a,rs99b} (RSII), the physical distance between the branes may
go off to infinity without affecting the validity of Newton's law on
our brane (due to additional massless graviton Kaluza-Klein modes).
\par

Interestingly, once one studies the cosmological evolution induced by
additional matter on the branes, the conventional (and successful)
relation $3H^2 = 8 \pi G_N \rho$ between the Hubble parameter $H$,
Newton's constant $G_N$ and the energy density $\rho$ is not always
obtained. Consequently the (non)-validity of this relation, at least at
temperatures below $\sim$~1~MeV, serves to eliminate certain scenarios.
\par

First, the cosmological evolution induced by matter on the branes has
been studied in the simplest case of vanishing cosmological constants
in the bulk and on the branes by Bin\'etruy, Deffayet and Langlois in
\cite{bdl99}. The astonishing result was that the conventional relation
$H^2 \sim \rho$ does not hold, rather one obtains $H \sim \rho$ (in
contradiction, e.g., with the otherwise successful big bang
nucleo-synthesis). \par

Subsequently, the matter-induced cosmological evolution in the RSI
scenario was investigated in \cite{cgkt99,cgs99a} with the ``negative''
result $H^2 \sim - \rho$ (if we live on the brane with negative
tension, which is a necessary condition for the solution of the
hierarchy problem in this case). \par

On the other hand, the cosmological evolution induced by matter on a
brane with positive tension agrees with the conventional evolution (at
late times) 
[17~--~23] if
one assumes that the physical distance between the branes is
stabilized, and the relative values of the cosmological constants in
the bulk and on the branes are fine-tuned. \par

The purpose of the present paper is the study, in the (compactified)
Ho{\v r}ava-Witten theory, of the cosmological evolution induced by
matter on the branes. We allow for an arbitrary equation of state,
i.e., a relation $p = w\rho$ among the pressure $p$ and the energy
density $\rho$. As fields in the bulk we consider, apart from the
components of the graviton, a scalar field $\varphi$ which corresponds
to the universal Calabi-Yau modulus. The presence of this field is
model independent and, due to the non-vanishing ``internal'' components
of the field strength associated to the 3-form, one obtains potentials
$V(\varphi )$ both in the bulk and on the branes
[4 -- 7]. If one introduces a dimensionful
parameter $\gamma$ with $\gamma \sim {\cal O} (M_{GUT}^4/M_{11}^3)$,
where $M_{11}$ is scale of the gravitational coupling in eleven
dimensions, one finds for the potential in the bulk $V^{(bulk)}(\varphi
) \sim {\cal O}(\gamma^2)$, whereas the potentials $V^{(n)}$ on the
branes satisfy $V^{(n)}(\varphi ) \sim {\cal O}(\gamma )$.  \par

Subsequently we will assume $\rho , p \ll \gamma \ \kappa_5^{-2}$ where
$\kappa_5^2$ is the inverse gravitational coupling in five dimensions
with mass dimension $\kappa_5^2 = M_5^{-3}$. This assumption is
certainly realistic with respect to ``Standard Model'' matter on our
brane, but our subsequent results will also apply to ``non-standard''
matter (e.g. associated with gaugino condensation on the hidden brane)
as long as the above inequality is satisfied. Notably we will allow for
an arbitrary dependence of the ``matter'' Lagrangians on the branes,
${\cal L}^{(n)}$, on the field $\varphi$: ${\cal L}^{(n)} = {\cal
L}^{(n)} (\varphi )$ (with ${\cal L}^{(n)} \sim {\cal O}(\rho , p) \ll
{\cal O}(\gamma \kappa_5^{-2}))$ which affects the junction conditions
of the field $\varphi$ on the branes, see below. \par

The inflationary evolution in this set-up, induced by matter potentials
on the branes (corresponding to an equation of state $p = \rho$), has
previously been discussed in \cite{low99}. The results obtained in 
\cite{low99} correspond either to the case $\gamma \cdot R_5 \ll 1$, 
which allows for a linearized approximation of the $y$ dependence of 
the fields, or to ``matter'' Lagrangians ${\cal L}^{(n)}$ on the branes
which are $\varphi$ independent. \par

Below, we will construct solutions in the bulk which are exact in $y$,
since we will \underbar{not} assume $\gamma \cdot R_5 \ll 1$. Notably,
we will not assume that the physical distance between the branes (or
the $yy$-component of the metric in the bulk) is stabilized in an ad
hoc fashion (corresponding mechanisms are discussed in 
[23,  25 -- 30]): One
of our aims is to see, under which general circumstances solutions with
a static size of the fifth dimension exist. In fact we will obtain,
depending on the $\varphi$ dependence of ${\cal L}^{(n)}(\varphi )$,
both solutions corresponding to a ``rolling radius'', as well as
solutions with a static $yy$-component of the metric. Not
astonishingly, only the latter case allows for a conventional
cosmological evolution on a brane. \par

In the next section we will present the general set-up of our approach:
the action in the bulk and on the branes, the Einstein equations and
the equation of motion of $\varphi$, the junction conditions on the
branes, and the exact (static) solution of \cite{losw98a}. In section 3
we introduce our ansatz for the time-dependent fields in the bulk,
which is motivated by the search for a conventional cosmological
evolution on the branes: we will assume, that the non-trivial
time-dependence is induced by the presence of matter on the branes. A
self-consistent ansatz is seen to be $\partial_t \sim {\cal O}((\rho
\gamma \kappa_5^2)^{1/2})$, where $\partial_t$ denotes the time
derivative of any field in the bulk or on the branes. (Given the
existence of time dependent solutions with $\partial_t \sim {\cal
O}(\gamma )$ \cite{low98,r98}, this ansatz is certainly not the most
general one. It represents, on the other hand, the ``minimal'' time
dependence which is induced by the presence of matter on the branes).
Then the Einstein equations and the equation of motion of $\varphi$ can
be solved exactly in $y$ and $t$, but neglecting terms of relative
order $\rho \kappa_5^2 \gamma^{-1}$. In section 4 we discuss the
physical properties of our solutions and conclude.  

\mysection{Equations of motion and junction conditions}
\hspace*{\parindent} 

Our starting point is a five dimensional action in the bulk which
depends, apart from the gravitational sector, on a scalar field
$\varphi$ which is related to the universal modulus of the internal
Calabi-Yau manifold 
[4 -- 7]. In the
notation of \cite{losw98a}, our field $\varphi$ is related to the field
$V$ in \cite{losw98a} by $V = \exp (2 \varphi )$. Likewise, our
parameter $\gamma$ appearing in the potentials of $\varphi$ in the bulk
and on the branes is related to the parameter $\alpha$ in
\cite{losw98a} through $\gamma = \sqrt{2} \alpha$. In any case $\gamma$
is of ${\cal O}(M_{GUT}^4/M_{11}^3)$. The exact value of $\gamma$
depends on the shape of the Calabi-Yau manifold \cite{w96,losw98a}. The
bulk action then reads 

\beq 
\label{2.1e}
S^{bulk} = - {1 \over \kappa_5^2} \int \sqrt{-g_5} \ d^5x \left \{ {1
\over 2} R + \partial_{\mu} \varphi \partial^{\mu} \varphi + {\gamma^2
\over 12} e^{-4 \varphi} \right \} \ .
\eeq

\noi Subsequently we denote the fifth dimension (across the bulk) by
$y$ and assume, that all fields depend on $t$ and $y$, but not on the 3
spatial coordinates $x_i$. A convenient (diagonal) ansatz for the
metric is then

\beq
\label{2.2e}
ds^2 = - e^{2\nu} dt^2 + e^{2\alpha} (d\vec{x})^2 + e^{2\beta} dy^2 \ .
\eeq

\noi Denoting time derivatives by points, and derivatives with respect
to $y$ by primes, the Einstein equations in the bulk become 

\bminiG{2.3e}
\label{2.3ae}
-e^{-2\nu} \left ( \dot{\alpha}^2 + \dot{\alpha}\dot{\beta} \right ) +
e^{-2\beta} \left ( \alpha'' + 2 \alpha '^2 - \alpha ' \beta ' \right )
= {\kappa_5^2 \over 3} \ T_0^0 \ , 
\eeeq  
\beeq \label{2.3be}
&&-e^{-2\nu} \left ( 2 \ddot{\alpha} + 3 \dot{\alpha}^2 - 2
\dot{\alpha}\dot{\nu} + \dot{\beta} (2 \dot{\alpha} - \dot{\nu}) +
\ddot{\beta} + \dot{\beta}^2 \right ) \nn \\  &&+ e^{-2 \beta} \left (
2 \alpha '' + \nu '' + 3 \alpha '^2 + \nu '^2 + 2 \alpha ' \nu ' -
\beta ' (2 \alpha ' + \nu ') \right ) = \kappa_5^2 \ T_S \ ,   
\eeeq 
\beeq \label{2.3ce} 
\dot{\alpha} \nu ' - \dot{\alpha}' - \dot{\alpha}
\alpha ' + \alpha ' \dot{\beta} = {\kappa_5^2 \over 3} \ T_{05} \ , 
\eeeq  
\beeq \label{2.3de} 
- e^{-2\nu} \left ( \ddot{\alpha} + 2
\dot{\alpha}^2 - \dot{\alpha} \dot{\nu} \right ) + e^{-2 \beta}
(\alpha'^2 + \alpha '\nu ') = {\kappa_5^2 \over 3} \ T_5^5 \ .
 \emini

\noi Here $T_S$ denotes the diagonal element of the spatial components
of the energy momentum tensor,

\beq
\label{2.4e}
T_i^j = T_S \delta_i^j \ .
\eeq

Given the action (\ref{2.1e}), and recalling that $\varphi$ depends
only on $t$ and $y$, the non-vanishing components of the energy
momentum tensor in the bulk (away from the branes) are

\bminiG{2.5e}
\label{2.5ae}
\kappa_5^2 \ T_0^0 = - e^{-2\nu} \dot{\varphi}^2 - e^{-2 \beta} \varphi
'^2 - {{\gamma^2}\over {12}} \ e^{-4 \varphi} \ , \eeeq  
\beeq
\kappa_5^2 \ T_S = e^{-2\nu} \dot{\varphi}^2 - e^{-2 \beta} \varphi
'^2 - {{\gamma^2}\over {12}} \ e^{-4 \varphi} \ , 
 \label{2.5be}
\eeeq 
\beeq
 \label{2.5ce}
\kappa_5^2 \ T_5^5 = e^{-2\nu} \dot{\varphi}^2 + e^{-2 \beta} \varphi
'^2 - {{\gamma^2}\over {12}} \ e^{-4 \varphi} \ , 
\eeeq 
\beeq
 \label{2.5de}
\kappa_5^2 \ T_{05} = 2 \dot{\varphi} \varphi ' \ .
\emini

\noi Finally we have to consider the equation of motion of the field
$\varphi$, which takes the form (away from the branes)

\bea
\label{2.6e}
&&- e^{-2 \nu} \left ( \ddot{\varphi} + \dot{\varphi} (3 \dot{\alpha} -
\dot{\nu} + \dot{\beta}) \right ) + e^{-2 \beta } \left ( \varphi '' +
\varphi ' (3 \alpha ' + \nu ' - \beta ' ) \right ) = - {\gamma^2 \over
6} \ e^{-4 \varphi} \ . 
\nn \\ \eea

Now we consider the actions on the branes, which are situated at
$y^{(n)} = 0, \pi R_5$. We parametrize them as

\beq
\label{2.7e}
S^{(n)} = \int \sqrt{-g_4} \ d^4x \left \{ {\cal L}_{matter}^{(n)}
(\varphi ) - V^{(n)} (\varphi ) \right \} \ . \eeq

 Here ${\cal L}_{matter}^{(n)}(\varphi )$ depends, of course, on many
additional matter fields $\phi_i$. Below, however, only the possible
dependence on the bulk field $\varphi$ will play a role. De\-pen\-ding
on the precise form of ${\cal L}_{matter}^{(n)}$, and on the
configurations of the fields $\phi_i$, this part of the brane actions
will contribute to the energy momentum tensors on the branes in the
form of energy densities $\rho^{(n)}$ and pressures $p^{(n)}$, see
below. \par

The potentials $V^{(n)}(\varphi )$ on the two branes are known to be of
the form \cite{losw98a,elpp98}

\beq \label{2.8e} 
V^{(n)} (\varphi ) = \mp {\gamma \over \kappa_5^2} \
e^{-2 \varphi} \ . \eeq

Here the minus sign applies to the brane at $y = 0$, and the plus sign
to the brane at $y = \pi R_5$. Since we assume 

\beq \label{2.9e} 
{\cal L}_{matter}^{(n)} \sim \rho^{(n)} \sim p^{(n)}
\ll {\gamma \over \kappa_5^2} \ ,  \eeq

\noi the brane at $y = 0$ is thus the brane with negative tension, if
$\gamma$ is positive. We can easily interchange the role of the two
branes (negative vs. positive tension) by changing the sign of
$\gamma$. \par 

As is well known, the presence of branes leads to additional singular
terms (proportional to $\delta$-functions in $y$) on the right-hand
sides of the Einstein equations (\ref{2.3ae}), (\ref{2.3be}), and the
equation of motion (\ref{2.6e}), which have to be matched by
singularities in the second derivatives in $y$ on the left-hand side.
Since all fields under con\-si\-de\-ra\-tion are symmetric under the
orbifold symmetry $Z_2$, these jumps in the first derivatives in $y$
fix these first derivatives completely at $y^{(n)} = 0, \pi R_5$. Here,
these junction conditions read

\bminiG{2.10e}
\label{2.10ae}
\alpha '^{(n)} = {\kappa_5^2 \over 6} \ e^{\beta} \ T_0^{0(n)} \ ,
\eeeq  \beeq
 \label{2.10be}
\nu '^{(n)} = {\kappa_5^2 \over 6} \ e^{\beta} \left ( 3 T_S^{(n)} -
2T_0^{0(n)} \right ) \ ,
\eeeq 
\beeq
 \label{2.10ce}
\varphi '^{(n)} = {\kappa_5^2 \over 4} \ e^{\beta} {\delta \over \delta
\varphi } \left ( V^{(n)}(\varphi ) - {\cal L}_{matter}^{(n)} (\varphi
) \right ) \ ,  \emini

\noi with

\bminiG{2.11e}
\label{2.11ae}
T_0^{0(n)} = - V^{(n)}(\varphi ) - \rho^{(n)} \ ,
\eeeq  \beeq
 \label{2.11be}
T_S^{(n)} = - V^{(n)}(\varphi ) + p^{(n)}
\emini

\noi and $V^{(n)}(\varphi )$ as in eq. (\ref{2.8e}) As discussed above,
$\rho^{(n)}$ and $p^{(n)}$ originate from the matter Lagrangian on the
branes, which we will not specify any further at this stage. \par

Static solutions to the Einstein equations (\ref{2.3e}), the scalar
equation of motion (\ref{2.6e}) and the junction conditions
(\ref{2.10e}) - in the absence of matter on the branes - have been
given in \cite{losw98a}. We will denote these solutions by
$\widetilde{\alpha}$, $\widetilde{\nu}$, $\widetilde{\beta}$ and
$\widetilde{\varphi}$, which read

\bminiG{2.12e} \label{2.12ae} 
\widetilde{\alpha}(y) =
\widetilde{\nu}(y) = {1 \over 2} \ell n \ H + A \ , 
\eeeq  
\beeq \label{2.12be} 
\widetilde{\beta} (y) = 2 \ell n \ H + B \ , 
\eeeq 
\beeq \label{2.12ce} 
\widetilde{\varphi}(y) = {3 \over 2} \ell n \ H + {1 \over 2} B \ , 
\eeeq  
\beeq \label{2.12de} 
H = {\gamma \over 3} |y| + C \qquad  {\rm for} \ - \pi R_5 \leq y \leq \pi
R_5 \ , \qquad H\left ( y + 2 \pi R_5 \right ) = H(y) \ . \emini

$A$, $B$ and $C$ are arbitrary constants; in terms of a supergravity
Lagrangian in four dimensions $B$ and $C$ correspond to linear
combinations of the arbitrary vevs of the standard fields $Re(S)$ and
$Re(T)$ \cite{low99}. \par

Time dependent solutions have been obtained in \cite{low98,r98}, but in
the next sections we seek for cosmological solutions, where the time
dependence is induced by the presence of matter on the branes. Hence
the static solutions (\ref{2.12e}) will serve as a starting point.

\mysection{Cosmological evolution induced by matter on the branes} 

\hspace*{\parindent} In this section we will construct solutions for
$\alpha$, $\nu$, $\beta$ and $\varphi$ in the presence of matter on the
branes. Matter on the branes contributes to the junction conditions
(\ref{2.10e}) which affects, by continuity in $y$, the solutions across
the bulk for all $y$. We will assume that the ``amount'' of matter is
small compared to the fundamental scales $M_{11}$, $M_{GUT}$ (which are
quite close to each other) and $M_5$, i.e. 

\beq \label{3.1e} \rho^{(n)} \sim p^{(n)} \sim {\cal
L}_{matter}^{(n)} \ll \gamma \kappa_5^{-2} \ . \eeq

On the other hand we will need no assumption on $\gamma R_5$, since we
will obtain solutions exact in $y$. \par

First we recall that, in the case of an ``empty'' bulk, the Hubble
constant $H$ was shown to satisfy $H \sim \rho$ \cite{bdl99}, i.e. time
derivatives $\partial_t$ are of the order $\partial_t \sim {\cal
O}(\rho )$. A corresponding result was also obtained in \cite{low99} in
the case where $\rho$ represents a $\varphi$ independent potential and
$\rho R_5 \kappa^2_5 \gg 1$.
Conventional cosmology, on the other hand, corresponds to
$H^2 \sim \rho$, i.e. $\partial_t \sim {\cal O}(\sqrt{\rho})$. This
latter behaviour can be obtained in the case of additional cosmological
constants in the bulk and on the branes
[17 --  23], and we will
also obtain corresponding solutions in the present case. \par

Our ansatz for solutions in the presence of matter on the branes will
be ge\-ne\-ra\-li\-za\-tions of the static solutions (\ref{2.12e}) in
two respects: First, we add $y$ and $t$ dependent contributions to
$\widetilde{\alpha}$, $\widetilde{\nu}$, $\widetilde{\beta}$ and
$\widetilde{\varphi}$, which are of ${\cal O}(\rho \kappa_5^2
\gamma^{-1})$ where $\rho$ denotes the order of magnitude of all terms
on the left-hand side of eq. (\ref{3.1e}). Second, we promote the
constants $A$, $B$ and $C$ in (\ref{2.12e}) to time dependent
parameters assuming, however, that all time derivatives are of the
order

\beq
\label{3.2e}
\partial_t \sim {\cal O}\left ( (\rho \gamma \kappa_5^2)^{1/2} \right )
\eeq

\noi or smaller. (This ansatz allows, of course, for a still ``weaker'' time dependence with
$\partial_t \sim {\cal O}(\rho )$). Thus we write

\bminiG{3.3e}
\label{3.3ae}
\alpha (y, t) = {1 \over 2} \ell n \ H (y, t) + \alpha_0 (t) +
\bar{\alpha} (y, t) \ ,
\eeeq  \beeq
 \label{3.3be}
\nu (y, t) = {1 \over 2} \ell n \ H (y, t) + \nu_0 (t) + \bar{\nu} (y,
t) \ ,
\eeeq 
\beeq
 \label{3.3ce}
\beta (y, t) = 2 \ell n \ H (y, t) + \beta_0 (t) + \bar{\beta} (y, t) \
,
\eeeq 
\beeq
 \label{3.3de}
\varphi (y, t) = {3 \over 2} \ell n \ H (y, t) + \varphi_0 (t) +
\bar{\varphi} (y, t) \ ,
\eeeq 
\beeq
 \label{3.3ee}
\beta_0(t) = 2 \varphi_0(t) \ ,
\eeeq 
\beeq
 \label{3.3fe}
H(y,t) = {\gamma \over 3} |y| + C(t) \ , 
\emini

\noi with

\beq
\label{3.4e}
\bar{\alpha}, \bar{\nu}, \bar{\beta} , \bar{\varphi} \sim {\cal O}\left
( \rho \kappa_5^2 \gamma^{-1} \right ) \ . \eeq

\noi Equation (\ref{3.3ee}) can be obtained from the dominant terms of
the Einstein equations, and agrees with the static limit (\ref{2.12e})
after $B \to 2\varphi_0(t)$. \par

Let us first use this ansatz in the (05) component of the Einstein
equations (\ref{2.3ce}). Subsequently we replace $\beta_0$ by $2
\varphi_0$ everywhere according to eq. (\ref{3.3ee}), which simplifies
several expressions. Using (\ref{3.2e}), (\ref{3.4e}), and keeping all
terms up to ${\cal O} (\rho^{3/2})$, eq. (\ref{2.3ce}) can be brought
into the form (with (\ref{2.5de}) for $T_{05}$)

\bea
\label{3.5e}
&&\left ( \dot{\alpha}_0 + {\dot{C} \over 2H} \right ) \left (
\bar{\nu}' - \bar{\alpha}'\right )  - \dot{\bar{\alpha}}' + \left ( 2
\dot{\varphi}_0 + 3 {\dot{C} \over H} \right ) \left ( \bar{\alpha}' -
{\bar{\varphi}' \over 3} \right ) \nn \\
&&+ {\gamma \over 6H} \left ( \dot{\bar{\beta}} - 2 \dot{\bar{\varphi}}
\right ) + {\gamma \over 6H^2} \dot{C} = 0 \ .
 \eea

One notes that all terms are of ${\cal O}(\rho^{3/2})$, except for the
last term on the left-hand side, which is a priori (from (\ref{3.2e}))
of ${\cal O}(\rho^{1/2})$. Thus $\dot{C}$ is exceptionally at most of
${\cal O}(\rho^{3/2})$, and will not contribute to the dominant orders
in $\rho$ in the following equations. \par

Next we insert our ansatz into the remaining Einstein equations
(\ref{2.3e}) as well as (\ref{2.6e}). We use eqs. (\ref{2.5e}) for the
components of the energy momentum tensor, and expand each expression in
$\rho$ using (\ref{3.2e}) and (\ref{3.4e}). The dominant terms of
${\cal O}(\rho^0)$ cancel as they should, and subsequently we display
all terms of ${\cal O}(\rho )$. The subsequent equations follow from
eqs. (\ref{2.3ae}), (\ref{2.3be}), (\ref{2.3de}) and (\ref{2.6e}),
after moving all time derivatives to the left (and with $\beta_0 = 2
\varphi_0$): 

\bminiG{3.6e}
\label{3.6ae}
&&\dot{\alpha}_0^2 + 2 \dot{\alpha}_0 \dot{\varphi}_0 - {1 \over 3}
\dot{\varphi}_0^2 = H^{-3} \ e^{2 \nu_0 - 4 \varphi_0} \left (
\bar{\alpha}'' + {\gamma \over 6H} \left ( 2 \bar{\varphi}' -
\bar{\beta}' \right ) \right . \nn \\
&&\left . + {\gamma^2 \over 18H^2} \left ( \bar{\beta} - 2
\bar{\varphi} \right ) \right ) \ , 
 \eeeq  \beeq
 \label{3.6be}
&&2 \ddot{\alpha}_0 + 3 \dot{\alpha}_0^2 - 2 \dot{\nu}_0 \left (
\dot{\alpha}_0 + \dot{\varphi}_0 \right ) + 4 \dot{\alpha}_0
\dot{\varphi}_0 + 2 \ddot{\varphi}_0 + 5 \dot{\varphi}_0^2 \nn
\\
&&= H^{-3}\ e^{2\nu_0-4 \varphi_0} \left ( 2 \bar{\alpha}'' +
\bar{\nu}'' + {\gamma \over 2H} \left ( 2 \bar{\varphi}' - \bar{\beta}'
\right ) + {\gamma^2 \over 6H^2} \left ( \bar{\beta} - 2 \bar{\varphi}
\right ) \right ) \ ,  \eeeq  \beeq
 \label{3.6ce}
&&\ddot{\alpha}_0 + 2 \dot{\alpha}_0^2 - \dot{\alpha}_0 \dot{\nu}_0 +
{1 \over 3} \dot{\varphi}_0^2 = H^{-3} \ e^{2\nu_0 - 4 \varphi_0} \left
( {\gamma \over 6H} \left ( 3 \bar{\alpha}' + \bar{\nu}' - 2
\bar{\varphi}' \right ) \right . \nn \\
&& \left . + {\gamma^2 \over 18 H^2} \left ( \bar{\beta} - 2
\bar{\varphi} \right ) \right ) , 
\eeeq  \beeq
 \label{3.6de}
&&\ddot{\varphi}_0 + \dot{\varphi}_0 \left ( 3 \dot{\alpha}_0 -
\dot{\nu}_0 + 2 \dot{\varphi}_0 \right ) = H^{-3} \ e^{2\nu_0 - 4
\varphi_0} \left ( \bar{\varphi}'' + {\gamma \over 2H} \left ( 3
\bar{\alpha}' + \bar{\nu}' - \bar{\beta}' \right ) \right . \nn \\
&&\left .  + {\gamma^2 \over 3H^2} \left (
\bar{\beta} - 2 \bar{\varphi} \right ) \right ) \ .  
\emini

\noi These four equations fix the $y$ dependence of $\bar{\alpha}$,
$\bar{\nu}$, $\bar{\beta}$ and $\bar{\varphi}$:

\bminiG{3.7e}
\label{3.7ae}
\bar{\alpha}(y,t) = \widehat{\alpha}(t) H^5 + {1 \over 3} F(y) \ ,
\eeeq  \beeq
 \label{3.7be}
\bar{\nu}(y,t) = \widehat{\nu}(t) H^5 + {1 \over 3} F(y) \ ,
\eeeq 
\beeq
 \label{3.7ce}
\bar{\beta}(y,t) = \widehat{\beta} (t) H^5 + 2F(y) + {2H \over \gamma}
F'(y) \ ,
\eeeq 
\beeq
 \label{3.7de}
\bar{\varphi}(y,t) = \widehat{\varphi}(t) H^5 + F(y)
\emini

\noi with $H$ as in eq. (\ref{2.12de}) and $F(y)$ arbitrary. \par

The four time-dependent parameters $\widehat{\alpha}$, $\widehat{\nu}$,
$\widehat{\beta}$ and $\widehat{\varphi}$ are constrained by the three
junction conditions (\ref{2.10e}) in terms of the matter on the branes.
Plugging our ans\"atz (\ref{3.3e}) as well as (\ref{3.7e}) into eqs.
(\ref{2.10e}), one finds again that the leading terms of ${\cal
O}(\rho^0)$ cancel, and the terms of ${\cal O}(\rho )$ can be brought
into the form

\bminiG{3.8e}
\label{3.8ae}
10 \gamma \varepsilon^{(n)} \widehat{\alpha} = - \left ( H^{(n)} \right
)^{-2}\kappa_5^2 \ \rho^{(n)}\ e^{2\varphi_0} + \gamma (\widehat{\beta}
- 2 \widehat{\varphi} ) \ , \eeeq  \beeq
 \label{3.8be}
10 \gamma \varepsilon^{(n)} \widehat{\nu} = \left ( H^{(n)} \right
)^{-2} \kappa_5^2 \left ( 3p^{(n)} + 2 \rho^{(n)} \right ) 
e^{2\varphi_0} + \gamma (\widehat{\beta} - 2 \widehat{\varphi} ) \ ,
\eeeq 
\beeq
 \label{3.8ce}
10 \gamma \varepsilon^{(n)} \widehat{\varphi} = - {3 \over 2} \left (
H^{(n)} \right )^{-2} \kappa_5^2  \ {\delta {\cal L}_{matter}^{(n)}
\over \delta \varphi} \  e^{2\varphi_0} + 3\gamma (\widehat{\beta} - 2
\widehat{\varphi} ) \ ,
\emini

\noi where $H^{(n)}$ denotes $H(y^{(n)})$ with $H(y)$ as in eq.
(\ref{2.12de}), and $\varepsilon^{(n)} = + 1$ for $y^{(n)} = 0$,
$\varepsilon^{(n)} = - 1$ for $y^{(n)} = \pi R_5$. The dependence on
the function $F(y)$ of eqs. (\ref{3.7e}) cancels in the junction
conditions (\ref{2.10e}). \par

Altogether we thus have six equations, three junction conditions at the
brane $n = 1$, and three junction conditions at the brane $n = 2$. Let
us first discuss, to what extent these six equations restrict the
properties of the matter on the different branes. We recall that, e.g.,
in the case of an ``empty'' bulk considered in \cite{bdl99},
$\rho^{(2)}$ is fixed in terms of $\rho^{(1)}$, and $p^{(2)}$ in terms
of $p^{(1)}$. In the present case the three equations which involve the
properties of the matter on the brane 2 can be brought into the form

\bminiG{3.9e}
\label{3.9ae}
 \left ( H^{(1)} \right )^2 p^{(1)} + \left ( H^{(2)} \right )^2
p^{(2)} = - \left ( H^{(1)} \right )^2 \rho^{(1)} - \left ( H^{(2)}
\right )^2 \rho^{(2)} \ , \eeeq  \beeq
 \label{3.9be}
 \left ( H^{(1)} \right )^2 {\delta {\cal L}_{matter}^{(1)} \over
\delta \varphi} + \left ( H^{(2)}\right )^2 {\delta {\cal
L}_{matter}^{(2)} \over \delta \varphi} = 2 \left ( H^{(1)} \right )^2
\rho^{(1)} + 2 \left ( H^{(2)} \right )^2 \rho^{(2)}\ , \eeeq  \beeq
 \label{3.9ce}
\gamma \kappa_5^2 \left [ \left ( H^{(1)} \right )^2 \rho^{(1)} + \left
( H^{(2)}\right )^2 \rho^{(2)} \right ] = 2 \left ( \widehat{\beta} - 2
\widehat{\varphi} \right ) \ .
 \emini

At this stage we have to recall that $H$, defined in eq.
(\ref{2.12de}), depends on an as yet arbitrary parameter $C$ (with
negligible time dependence): we have $H^{(1)} = C$, $H^{(2)} = {\gamma
\over 3} \pi R_5 + C$. Hence eqs. (\ref{3.9e}) do not necessarily
constrain the matter on the brane 2 in terms of the matter on the brane
1, but can rather serve to fix $C$. \par

Of particular interest is the case where the matter fields $\phi_i$ on
the branes are constant (at the minima of their potentials), and the
only role of ${\cal L}_{matter}^{(n)}$ is thus to provide additional
(possibly constant) potentials for the modulus field $\varphi$ on the
branes. Then we can write

\bminiG{3.10e}
\label{3.10ae}
{\cal L}_{matter}^{(n)} = - \widehat{V}^{(n)} (\varphi ) \ ,
\eeeq  \beeq
 \label{3.10be}
\rho^{(n)} = \widehat{V}^{(n)}(\varphi ) \ ,
\eeeq 
\beeq
p^{(n)} = - \widehat{V}^{(n)}(\varphi) \ .
 \label{3.10ce}
\emini

\noi Hence eq. (\ref{3.9ae}) is satisfied identically. Eq.
(\ref{3.9be}) serves to fix $C$ in terms of the potentials on the
branes and $\varphi$. However, in order not to contradict the previous
result $\dot C\ \lsim\ {\cal O}(\rho^{3/2})$, one must have either
$\dot{\varphi}_0 \sim 0$, or $\widehat{V}^{(1)}(\varphi ) = {\rm
const.} \ \widehat{V}^{(2)}(\varphi )$. Eq. (\ref{3.9ce}) fixes the
combination $(\widehat{\beta} - 2 \widehat{\varphi})$. \par

Herewith we conclude the discussion on the relation between the matter
on the different branes, and concentrate subsequently on the physics on
brane 1 assuming that eqs. (\ref{3.9e}) are satisfied. Note, however,
that the two branes are physically distinct in Ho{\v r}ava-Witten
theory, since the dominant potentials $V^{(n)}(\varphi )$ on the branes
differ in their sign (cf. eq. (\ref{2.8e})). On the other hand we find
from eqs. (\ref{3.8e}) that we can exchange the role of the two branes
by changing the sign of $\gamma$, and redefining the parameter
$\widehat{\beta}$:

\beq
\label{3.11e}
\gamma \to - \gamma \ , \ \widehat{\beta} \to 4 \widehat{\varphi} -
\widehat{\beta} \ . \eeq

In order to solve eqs. (\ref{3.8e}) on brane 1 it is convenient to
define the parameters $w$, $d$ as follows:  

\bminiG{3.12e}
\label{3.12ae}
p^{(1)} = w \rho^{(1)} \ ,
\eeeq  \beeq
 \label{3.12be}
{\delta {\cal L}_{matter}^{(1)} \over \delta \varphi} = - d \rho^{(1)}
\ .
\emini

\noi In general the parameter $d$ will depend on $\varphi$, unless
${\cal L}_{matter}^{(1)}$ and $\rho^{(1)}$, $p^{(1)}$ happen to be
related as, e.g., in eqs. (\ref{3.10e}) with

\beq
\label{3.13e}
\widehat{V}^{(n)} (\varphi ) \sim e^{d\varphi} \ .
\eeq

\noi With eqs. (\ref{3.12e}) and $H^{(1)} = C$, the general solution of
eqs. (\ref{3.8e}) on brane 1 can be written as

\bminiG{3.14e}
\label{3.14ae}
\widehat{\alpha} = {1 \over 16} \ \widehat{\beta} - {\kappa_5^2 \
\rho^{(1)} \ e^{2 \varphi_0} \over 160 \gamma C^2} \ (3d + 16) \ ,
\eeeq  \beeq
 \label{3.14be}
\widehat{\nu} = {1 \over 16} \ \widehat{\beta} + {\kappa_5^2 \
\rho^{(1)} \ e^{2\varphi_0} \over 160 \gamma C^2} \left ( - 3d + 16 (3w
+ 2) \right ) \ , \eeeq 
\beeq
 \label{3.14ce}
\widehat{\varphi} = {3 \over 16} \ \widehat{\beta} + {3 \kappa_5^2 \
\rho^{(1)} \ e^{2\varphi_0} \over 32 \gamma C^2} \ d \ ,
 \emini

\noi with $\widehat{\beta}$ arbitrary. (Eventually $\widehat{\beta}$
can be fixed, combining eqs. (\ref{3.14ce}) and (\ref{3.9ce}), in terms
of $\rho^{(2)}$). \par

Combining eqs. (\ref{3.14e}) and eqs. (\ref{3.7e}) we have thus
obtained the general solutions for $\bar{\alpha}$, $\bar{\nu}$,
$\bar{\beta}$, $\bar{\varphi}$ in terms of the properties of the matter
on the branes and for all values of $y$. \par

Let us recall the ansatz (\ref{3.3e}) for the fields $\alpha$, $\nu$,
$\beta$ and $\varphi$. Since we obtained $\dot{C}\ \lsim\ {\cal O}
(\rho^{3/2})$, the dominant contributions to these fields decompose
into a sum of $y$ dependent and $t$ dependent terms. The cosmological
evolution on any brane is thus determined by $\alpha_0(t)$, $\nu_0(t)$
and $\varphi_0(t)$. The knowledge of the subdominant contributions
$\bar{\alpha}$, $\bar{\nu}$, $\bar{\beta}$ and $\bar{\varphi}$ to eqs.
(\ref{3.3e}) is required in order to obtain the $t$ dependence of
$\alpha_0$, $\nu_0$ and $\varphi_0$ from eqs. (\ref{3.6e}), which we
exploit in the following. \par

Once we plug our solutions (\ref{3.7e}) for $\bar{\alpha}$,
$\bar{\nu}$, $\bar{\beta}$ and $\bar{\varphi}$, together with eqs.
(\ref{3.14e}), into eqs. (\ref{3.6e}), we first observe that all
dependence on the arbitrary function $F(y)$ in (\ref{3.7e}) as well as
on $\widehat{\beta}$ cancels out. Hence the time dependence of
$\alpha_0$, $\nu_0$ and $\varphi_0$ can be related to the properties of
the matter on the brane 1 only, which are encoded in our parameters
$\rho^{(1)}$, $w$ and $d$. Second, we may choose the gauge $\nu_0(t) =
0$, such that $t$ is proportional - up to small corrections - to the
cosmic time: now we have $\partial_t \nu (y, t)\ \lsim\ {\cal
O}(\rho^{3/2})$ (cf. eq. \ref{3.3be}), since $\dot{C}$ and
$\dot{\bar{\nu}}$ are at most of ${\cal O} (\rho^{3/2})$. \par

Since one finds that one of eqs. (\ref{3.6e}), e.g. eq. (\ref{3.6be}),
is redundant, we just give our results for eqs. (\ref{3.6ae}),
(\ref{3.6ce}) and (\ref{3.6de}):

\bminiG{3.15e}
\label{3.15ae}
\dot{\alpha}_0^2 + 2 \dot{\alpha}_0 \dot{\varphi}_0 - {1 \over 3}
\dot{\varphi}_0^2 = - {2 \over 9} {\gamma \kappa_5^2 \rho^{(1)} \over
C^2} \ e^{-2 \varphi_0} \eeeq  \beeq
 \label{3.15be}
\ddot{\alpha}_0 + 2 \dot{\alpha}_0^2 + {1 \over 3} \dot{\varphi}_0^2 =
{\gamma \kappa_5^2 \rho^{(1)} \over 36 C^2} \ e^{-2 \varphi_0} (3w - 1
- 3d) \ , \eeeq 
\beeq
 \label{3.15ce}
\ddot{\varphi}_0 + \dot{\varphi}_0 \left ( 3 \dot{\alpha}_0 + 2
\dot{\varphi}_0 \right ) = {\gamma \kappa_5^2 \rho^{(1)} \over 12 C^2}
\ e^{-2 \varphi_0} (3w - 1 + d) \ . 
\emini

\noi Eqs. (\ref{3.15e}) justify, a posteriori, our initial ansatz
(\ref{3.2e}) for the order of $\partial_t$, provided $C^2 \exp (2
\varphi_0) \sim {\cal O}(1)$. \par

Taking the time derivative of eq. (\ref{3.15ae}) and using all eqs.
(\ref{3.15e}), the analog of the standard energy conservation condition
can be obtained:

\beq
\label{3.16e}
\dot{\rho}^{(1)} = - 3 \rho^{(1)} \ \dot{\alpha}_0 (1 + w) + d
\rho^{(1)} \dot{\varphi}_0 \ .
\eeq 

First, if we insert our solution (\ref{3.7e}) for $\bar{\alpha}$,
$\bar{\nu}$, $\bar{\beta}$ and $\bar{\varphi}$, together with eqs.
(\ref{3.14e}) and eq. (\ref{3.16e}), into eq. (\ref{3.5e}), we obtain a
trivial time dependence of $C$:

\beq
\label{3.17e}
\dot{C} = 0 \ .
\eeq

Second, eq. (\ref{3.16e}) differs from the standard condition for
energy conservation due to the last term. This term describes the
``disappearance'' of energy into the fifth dimension, if both $\delta
{\cal L}_{matter}^{(1)}/\delta \varphi \sim d\not= 0$ and
$\dot{\varphi}_0 \not= 0$. In fact, if ${\cal L}_{matter}^{(1)}$ is of
the form of a potential $-\widehat{V}^{(1)}(\varphi )$ (as in eqs.
(\ref{3.10e})) we have $w = -1$ and, using $\dot{\varphi} \sim
\dot{\varphi}_0$, eq. (\ref{3.10be}) and eq. (\ref{3.12be}),

\beq
\label{3.18e}
\dot{\rho}^{(1)} = \partial_t \ \widehat{V}^{(1)} (\varphi ) = {\delta
\widehat{V}^{(1)} \over \delta \varphi} \dot{\varphi} \cong -{\delta
{\cal L}_{matter}^{(1)} \over \delta \varphi} \dot{\varphi}_0 =
d\rho^{(1)} \dot{\varphi}_0 \eeq

\noi in agreement with (\ref{3.16e}). \par

For $d \not= 0$, a ``standard'' cosmological evolution on any brane,
which requires standard energy conservation, is only possible for
$\dot{\varphi}_0 = \dot{\beta}_0 = 0$. From eq. (\ref{3.15ce}) this
situation is feasable only for

\beq
\label{3.19e}
d = 1 - 3 w \ ,
\eeq  

\noi which requires a particular dependence of ${\cal
L}_{matter}^{(1)}$ on $\varphi$. If, again, ${\cal L}_{matter}^{(1)}$
is of the form of a potential (hence $w = - 1$), we need $d = 4$ or

\beq
\label{3.20e}
\widehat{V}^{(1)}(\varphi ) = {\rm const.} \ e^{4 \varphi} \ .
\eeq

\noi Furthermore, from eq. (\ref{3.15ae}), $\dot{\varphi}_0 = 0$
implies immediately

\beq
\label{3.21e}
\gamma \cdot \rho^{(1)} \leq 0 \ .
\eeq 

\noi Thus, if we want to accomodate a standard positive energy density
$\rho^{(1)}$, we have to choose $\gamma < 0$, i.e. brane 1 has to be
the one with a positive tension (since now the dominant contribution
$V^{(1)}(\varphi )$, from eq. (\ref{2.8e}), is positive). This result
coincides with the one obtained in the case of cosmological constants
in the bulk and on the branes
[17  -- 23]. \par

If eq. (\ref{3.19e}) is satisfied, eqs. (\ref{3.15e}) allow for
$\dot{\varphi}_0 = 0$. Since we have $\dot{\alpha} \sim \dot{\alpha}_0$
and, from our definition (\ref{2.2e}) of the metric, $\dot{\alpha}$
corresponds to the Hubble parameter, eqs. (\ref{3.15ae}) and
(\ref{3.15be}) are easily seen to correspond to the ordinary Freedman
equations for $\gamma < 0$. (A constant term in $\nu$ allows for a
constant rescaling of $t$ such that the right-hand sides of eqs.
(\ref{3.15ae}) and (\ref{3.15be}), for $\dot{\varphi}_0 = 0$, can
always be brought into standard form). \par

If eq. (\ref{3.19e}) is not satisfied, we have necessarily
$\dot{\varphi}_0 \not= 0$ and, from eq. (\ref{3.3ee}), $\dot{\beta}_0
\not= 0$, i.e. solutions corresponding to a ``rolling radius'': the
physical distance between the two branes, given by $\int e^{\beta
(y)}dy$ with $y = \{0, \pi R_5\}$, varies with $t$. \par

Explicit solutions to eqs. (\ref{3.15e}) can be given if $w =$ const.
and $d =$ const. From eq. (\ref{3.12be}) the second condition is
satisfied if

\beq
\label{3.22e}
{\cal L}_{matter}^{(1)} (\varphi ) \sim \rho^{(1)}(\varphi , t) =
\widehat{\rho}^{(1)} (t) e^{d\varphi} \ . \eeq

\noi Then eqs. (\ref{3.15e}) are solved for

\bea
\label{3.23e}
&&\alpha_0(t) = {\rm const.} + r \ln t \ , \ \varphi_0 (t) = {\rm
const.} + s \ln t \ , \nn \\
&&\rho^{(1)} \sim {\rm const.} \ t^{2s-2} \quad \left ( {\rm or} \
\widehat{\rho}^{(1)} \sim {\rm const.} \ t^{(2 -d) s-2}\right ) \ , 
\eea 

\noi with

\beq
\label{3.24e}
r = {2(w - 3 - d) \over 3w^2 - 6wd - d^2 - 11} \ , \ s = {2(3w - 1 + d)
\over 3w^2 - 6wd - d^2 - 11} \ ,  \eeq

\noi provided the denominators are non-zero. In the case where ${\cal
L}_{matter}^{(1)}(\varphi )$ is just an additional potential in
$\varphi$, as in eqs. (\ref{3.10e}) with $n = 1$, we have $w = - 1$ and
thus

\beq
\label{3.25e}
r = {2(4+d) \over (4 - d) (2 - d)} \ , \ s = {2 \over 2 - d} \quad {\rm
if} \ d\not= 2, 4 \ . \eeq

(The particular case $d = -2$, where $\widehat{V}^{(n)} (\varphi)$ in
eqs. (\ref{3.10e}) has the same functional dependence on $\varphi$ as 
${V}^{(n)} (\varphi)$ in eq. (\ref{2.8e}), has already been considered
in \cite{lid99}.) 
For $d = 4$ we are back in the situation where eq. (\ref{3.19e}) is
valid (and where standard inflation on the branes is obtained for $w =
- 1$), whereas for $d = 2$ we obtain inflationary evolution both on the
branes and across the bulk, i.e. in $\beta (t) \sim 2 \varphi_0 (t)$:

\bea
\label{3.26e}
&&a(t) = e^{\alpha (t)} \sim e^{\alpha_0 (t)} \sim e^{const. \ t} \ ,
\nn \\
&&\dot{\varphi}_0 = {1 \over 3} \dot{\alpha}_0 \ , \ \rho (t) =
\bar{\rho} \ e^{2 \varphi_0 (t)}
\eea 

\noi with $\dot{\bar{\rho}} = 0$, and const. $\sim \pm
\sqrt{\bar{\rho}}$. \par

Herewith we conclude the different solutions to eqs. (\ref{3.15e}),
which will be discussed in the next section.

\mysection{Discussion and conclusions} 

\hspace*{\parindent} In the previous section we have constructed
various cosmological solutions in (compactified) Ho{\v r}ava-Witten
theory with additional matter on the branes, as\-su\-ming $\rho \ll
\gamma \kappa_5^{-2}$, cf. eq. (\ref{3.1e}). Actually, whenever $\rho$
varies with $t$, this assumption restricts the validity of the
solutions to corresponding regimes in $t$, typically to sufficiently
late time once $\rho$ decays in $t$. \par

Generally compactified Ho{\v r}ava-Witten theory suffers from the usual
moduli problem, i.e. there are scalar degrees of freedom with vanishing
potentials. In the present framework this phenomenon corresponds to the
presence of arbitrary constants in the static solutions (\ref{2.12e})
\cite{losw98a} (in the absence of additional matter), and to the
existence of time dependent solutions with $\partial_t \sim {\cal
O}(\gamma )$ \cite{low98,r98}. Since we assumed a much weaker time
dependence, cf. eq. (\ref{3.2e}), our solutions require
par\-ti\-cu\-lar initial conditions in the form of nearly static (and
homogeneous) configurations. However, once matter on the branes is
present, the fields cannot vary slower with $t$ than indicated by our
solutions. \par

One aspect of the moduli problem is the fact that generically the
physical distance between the branes, i.e. the $yy$ component of the
metric (parametrized by $\beta$), varies in time. Our results show
that, once matter is present only on the branes, this phenomenon is
unavoidable, unless the matter couples to the universal Calabi-Yau
modulus field $\varphi$ in a particular way, cf. eq. (\ref{3.19e}).
Possibly the five dimensional radius can be stabilized by means of an
additional potential for $\varphi$ in the bulk
[23, 25 -- 30]. Then our
cosmological solutions would be relevant at times $t$, at which
potentials $V^{(n)}(\varphi ) \ll \gamma \kappa_5^{-2}$ on the branes
are present (e.g. due to gaugino condensation), but where additional
bulk-potentials are not yet switched on. \par

Let us summarize our results in the case where the additional matter
Lagrangian on the branes corresponds to potentials for $\varphi$, i.e.
where the equation of state corresponds to $w = - 1$. Exact solutions
have been obtained for potentials of the form

\beq
\label{4.1e}
\widehat{V}^{(n)}(\varphi ) \left ( = - \rho^{(n)} \right ) =
\widehat{\rho}^{(n)}\ e^{d\varphi} \ .
 \eeq

Generically we obtain solutions with a power law behaviour in $t$ for
the scale factor $a$ and the physical distance $R_{phys}$ between the
branes (cf. eqs. (\ref{3.23e}) and below; note that
$\widehat{\rho}^{(n)}$ is time independent for $w = - 1$, where eqs.
(\ref{3.25e}) are valid):

\bminiG{4.2e}
\label{4.2ae}
a(y, t) = e^{\alpha (y,t)} \sim {\rm const.}(y) \ e^{\alpha_0 (t)} \sim
t^r \ ,
\eeeq  \beeq
 \label{4.2be}
R_{phys}(t) \sim e^{\beta_0 (t)} = e^{2 \varphi_0 (t)} \sim t^{2s} \ ,
\emini  

\noi with $r$, $s$ as in eqs. (\ref{3.25e}) (for $d = 0$, e.g., one has
$r = s = 1$). In the particular case $d=2$ one obtains, from eqs.
(\ref{3.26e}), inflationary (i.e. exponentially increasing or
decreasing) evolutions for $a$, $R_{phys}$ and $\widehat{\rho}^{(n)}$.
In the other particular case $d= 4$ the ordinary Freedman equations
hold on the branes: Now $R_{phys}$ can be time independent (since
$\dot{\beta}_0 = 2 \dot{\varphi}_0 = 0)$, and inflationary evolution
happens only ``parallel'' to the branes in terms of an exponential $t$
dependence of $a$. \par

Once the equation of state of the matter on the branes differs from $w
= -1$ we can still obtain solutions with constant $R_{phys}$, if eq.
(\ref{3.19e}) is satisfied. In these cases the cosmological evolution
is of the standard form. For radiation dominated matter ($w = {1 \over
3}$), e.g., standard cosmological evolution is obtained for $d= 0$,
i.e. when the (dominant part of the) matter action is independent of
$\varphi$. For nonrelativistic matter ($w = 0$) standard cosmological
evolution is obtained once the matter action satifies eq.
(\ref{3.12be}) with $d= 1$. It remains to be seen whether these
solutions constitute an alternative to the (presently ad hoc) radius
fixation by potentials in the bulk, i.e. whether corresponding
couplings of the Calabi-Yau modulus $\varphi$ to matter on the branes
and sufficiently well-behaved initial conditions can be obtained. \par

In any case we have seen that matter induced cosmological evolution in
Ho{\v r}ava-Witten theory (albeit in its simplest version with just the
universal Calabi-Yau modulus in the bulk) differs considerably from
simpler scenarios as an empty bulk or a cosmological constant in the
bulk.

\end{document}